\documentclass[]{aa}
\usepackage{txfonts}
\usepackage {graphicx}
\usepackage{placeins}
\usepackage {natbib}
\bibpunct{(}{)}{;}{a}{}{,}
\usepackage{longtable,array,tabularx,afterpage,booktabs}
\usepackage[figuresright]{rotating}

\begin{document}
\title{Compact radio emission from $z\sim0.2$ X-ray bright AGN}
\author{J. Zuther\inst{1}, S. Fischer\inst{1}, A. Eckart\inst{1,2}}
\institute{I. Physikalisches Institut, Universit\"at zu K\"oln,
Z\"ulpicher Str. 77, 50937 K\"oln, Germany 
\and Max-Planck-Institut f\"ur Radioastronomie, Auf dem H\"ugel 69, 53121 Bonn, Germany}
\offprints{J. Zuther, \email{zuther@ph1.uni-koeln.de}}
\date{Received  / Accepted }
\abstract
{Radio and X-ray emission of AGN appears to be correlated. The details of the underlying physical processes, however, are still not fully understood, i.e., to what extent is the X-ray and radio emission originating from the same relativistic particles or from the accretion-disk or corona or both.}
{We study the cm radio emission of an SDSS/ROSAT/FIRST matched sample of 13 X-raying AGN in the redshift range $0.11\leq z \leq 0.37$ at high angular resolution with the goal of searching for jet structures or diffuse, extended emission on sub-kpc scales.}
{We use MERLIN at 18~cm for all objects and Western EVN at 18~cm for four objects to study the radio emission on scales of $\sim$500~pc and $\sim$40~pc for the MERLIN and EVN observations, respectively.}
{The detected emission is dominated by compact nuclear radio structures. We find no kpc collimated jet structures. The EVN data indicate for compact nuclei on 40~pc scales, with brightness temperatures typical for accretion-disk scenarios.
Comparison with FIRST shows that the 18~cm emission is resolved out up to 50\% by MERLIN. 
Star-formation rates based on large aperture SDSS spectra are generally too small to produce considerable contamination of the nuclear radio emission. 
We can, therefore, assume the 18~cm flux densities to be produced in the nuclei of the AGN. 
Together with the ROSAT soft X-ray luminosities and black hole mass estimates from the literature, our sample objects follow closely the \citeauthor{2003MNRAS.345.1057M} fundamental plane relation, which appears to trace the accretion processes. Detailed X-ray spectral modeling from deeper hard X-ray observations and higher angular resolution at radio wavelengths are required to further proceed in the disentangling of jet and accretion related processes.}
{}


\authorrunning{J. Zuther et al.}
\titlerunning{X-ray bright AGN at 18 cm}
\maketitle
\section{Introduction}

X-rays have turned out to be efficient tracers of nuclear activity in active galactic nuclei (AGN) \citep{2005ARA&A..43..827B}. Current models suggest that the observed X-rays are produced in the accretion flow onto the central supermassive black hole (SMBH), with soft X-rays typically originating from the accretion disk and hard, comptonized X-rays from a corona surrounding the accretion disk \citep[e.g.,][]{1978Natur.272..706S,1991ApJ...380L..51H,2002A&A...389..802P}.

Related to the accretion phenomenon is the observation of jets, outflows which consist of charged particles (primarily electrons) accelerated to relativistic velocities.
The electrons give rise to the observed synchrotron spectrum at radio frequencies \citep[e.g.,][]{1984ARA&A..22..319B}. Some AGN, however, show flat radio spectra, which is interpreted as self-absorbed synchrotron radiation of multiple jet knots \citep[e.g.,][]{1980ApJ...238L.123C}.

Several studies of radio and X-ray detected AGN have resulted in an apparent correlation between the radio and X-ray luminosities \citep[e.g.,][]{2004ApJ...600L..31S,2000A&A...356..445B}, suggesting a common physical ground of both phenomena. Together with the associated black-hole mass, these three values determine the location of objects in the fundamental plane of black hole activity \citep[FP; e.g.,][]{2004A&A...414..895F,2003MNRAS.345.1057M}. 
It is, however, still not clear what the physical relation between the three properties is. 
Several structural components (accretion disk, corona, jet, disk wind) and physical processes (synchrotron, (inverse) Compton, bremsstrahlung, relativistic beaming) can contribute to the total radio and X-ray output \citep[e.g.,][]{1998tbha.conf..148N, 2004MNRAS.355L...1H, 2005ApJ...620..905Y, 2011MNRAS.413.1735S}. 
Their relative importance of which appears to depend on the accretion efficiency, the black hole mass, and potentially the black hole spin \citep[e.g.,][]{2006A&A...456..439K, 2007ApJ...658..815S}. 
This, however, renders a simple interpretation of the FP, i.e. identification of physical processes by locus in the FP, impractical, at least for individual objects. 
Rather, there appears to be a range of different FPs, depending on object class, i.e. the predominant physical processes.  
Detailed analyses are, therefore, necessary, in order to properly disentangle star formation, accretion flow, and jet emission from each other in the circumnuclear environment of AGN.
%
	This is especially important at higher redshifts, for which the probed linear scales become increasingly larger and the surface brightnesses increasingly dimmer.  
Sensitive, high angular resolution observations at radio wavelengths are helpful in this respect, as they allow to assess the presence and importance of jet emission, relativistic beaming, and star formation, based on radio morphology and radio spectral index.
	For example, Japanese VLBI Network (JVN) milli arc-second angular 	resolution ($\sim 4$~mas) observations of three $z\sim 0.2$ RL NLS1 galaxies \citep{2007PASJ...59..703D} show unresolved nuclei on pc scales and brightness temperatures, as well as, spectral properties indicative for relativistically beamed emission. 
	\cite{2005ApJ...621..123U} carried out VLBA imaging of RQ QSOs at mas resolution. Two objects of their sample are at $z\approx 0.2$ and either show a jet or a self-absorbed, compact synchrotron source. According to the ratio of their X-ray and radio fluxes, they appear to be rather radio loud \citep[cf.,][]{2003ApJ...583..145T}.
	Based on VLBA observations, \cite{1998MNRAS.299..165B} find strong evidence for jet-producing central engines for some of their RQ QSOs, in the form of relativistically beamed radio emission. 
	These objects show relativistically beamed emission. 
	\cite{2006A&A...455..161L} present VLA configuration A and B imaging of $z\approx 0.2$ RQ QSOs and low-$z$ Seyfert galaxies. The radio structures found in such QSOs resemble those of jets and outflows, and the authors propose RQ QSOs to be up-scaled versions of Seyfert galaxies.
	A question, however, remains, why these objects cannot form the powerful jets as have been found in RL QSOs, although they possess the jet-producing central engine.
	\cite{2011MNRAS.413.1735S} follow the idea that the radio emission of RQ QSOs is produced by free-free emission in the optically thin part of an accretion disc wind. Studying RQ Palomar Green QSOs (PG QSOs) in X-rays and in the radio at sub arc-second resolution, they found that only in two out of twenty objects, bremsstrahlung might contribute to the nuclear radio emission. 
	The QSOs investigated in the literature so far are among the brightest members of the intermediate redshift AGN population. 
	Studying the radio properties of less luminous QSOs, therefore, allows to probe and extend the parameter space of relevant physical processes to less powerful and more abundant QSOs.   

In the following, we present radio interferometric 18~cm continuum observations of 13 optically selected, X-ray bright AGN to further investigate the radio/X-ray correlation for intermediate redshift AGN. 
	Section \ref{sec:sample} describes the sample and Sect. \ref{sec:obs} the observations. Section \ref{sec:results} presents the results and a discussion in the context of the radio/X-ray correlation, concluded by a summary in Sect. \ref{sec:summary}. 
The Appendix provides notes on individual targets.
Throughout this work, we use a cosmology with $H_0=70$~km~s$^{-1}$~Mpc$^{-1}$, $\Omega_m=0.3$, and $\Omega_\Lambda=0.7$ \citep{2003ApJS..148..175S}. 

\section{The sample}
\label{sec:sample}
The space density of luminous AGN is very low in the nearby Universe \citep[e.g.,][]{2009A&A...507..781S}. In order to study the (circum-) nuclear energetics in such AGN, one has to resort to a larger cosmological volume. 
Increased projected linear scales and surface-brightness dimming, however, make more distant galaxies harder to interpret. A proper disentanglement of Seyfert and star-formation features benefits from radio-interferometric observations.
These are possible with current observatories like the Multi-Element Radio Linked Interferometer Network (MERLIN) or the European Very Long Baseline Interferometer Network (EVN). 
Observations of luminous, intermediate redshift AGN have the potential to impart means to connect the physical drivers in terms of fueling and feedback of low-redshift and less powerful AGN to those of high-redshift, very powerful AGN.
%

The 13 targets\footnote{The last four objects in Tab. \ref{tab:basicParam} have nearby ($<40\arcsec$), bright ($r<15$~mag) guide stars and are suitable for adaptive optics assisted follow-up observations of their host galaxies at near-IR wavelengths \cite[cf.,][]{2009mavo.proc...15Z}.} presented in this work are based on a cross-match between the Sloan Digital Sky Survey 2nd data release \citep[SDSS DR2,][]{2004AJ....128..502A} and the ROSAT All Sky Survey \citep[RASS,][]{1999A&A...349..389V}. From the SDSS, we have chosen only those objects, for which an optical spectrum has been measured. The SDSS/RASS pairs were subsequently matched with the VLA Faint Images of the Radio Sky at Twenty centimeters survey \citep[FIRST,][]{1995ApJ...450..559B}. FIRST provides better than 5\arcsec angular resolution.
SDSS, RASS, and FIRST are very well matched in terms of sky coverage and sensitivity \citep{2003AJ....126.2209A}.
For this pilot study, we specifically selected intermediate redshift objects ($0.1<z<0.4$) from the matched sample, which lie close to the classical Seyfert/QSO luminosity demarcation \citep[i.e., $M_i=-22$][]{2003AJ....126.2579S} and which encompass different spectral types of AGN (i.e., Seyfert 1, Narrow Line Seyfert 1, LINER, radio galaxy; cf., Tab. \ref{tab:basicParam} for individual classes).
These targets, for given redshift, are less luminous than the well-studied PG QSOs \citep[by about 2 mag at $V$ band; cf.,][]{2008MNRAS.390..847L}.
Furthermore, we selected only FIRST-unresolved RQ to radio-intermediate objects. They have milli-Jansky flux densities, making them accessible for sensitive radio interferometric observations
and allow us to investigate the nature of their core radio emission. 
The details of the X-ray/optical matching are described in \cite{2009mavo.proc...15Z}.
The matched sample, at this point, is by no means complete, because of the selection of specific AGN classes from the quite small initial crossmatched sample (about 100 objects). Resorting to more recent versions of crossmatched samples (e.g., \citealt{2007AJ....133..313A}; Vitale et al. 2012, submitted to A\&A; Zuther et al. in prep.; based on more recent SDSS data releases), allow for the selection of statistically more relevant subsamples of individual AGN classes (e.g., NLS1s) for high angular resolution follow-up studies.   

\begin{table*}[ht!]
\caption{Basic SDSS, ROSAT, and FIRST characteristics of the sample.}
\begin{center}
\begin{tabular}{c c c c c c c c c c c}
\hline
\hline
ID & Name (ID)& $z$& Class$^{\mathrm{a}}$& Env.$^{\mathrm{b}}$& M$_{i,\mathrm{PSF}}$ & $F_\mathrm{20~cm, peak}$& $\theta_\mathrm{20~cm}$$^{\mathrm{c}}$& $\log R*$ & $\log F_\mathrm{0.1-2~keV}$&$\log M_\bullet$$^{\mathrm{d}}$\\
&&&&&[mag]&[mJy]&[sqarcsec]&&[erg s$^{-1}$ cm$^{-2}$]&[$M_\odot$]\\
(1) & (2) & (3) & (4) & (5) & (6) & (7) & (8) & (9)&(10)&(11)\\
\hline
1                   & \object {SDSS J153911.17+002600.7} & 0.265 &NLS1    &Int &  -22.6&  $ 1.11\pm 0.14$& $ 3.9\times --   $ &0.6 & -12.4 & 6.5 \\
2                   & \object {SDSS J150521.92+014149.8} & 0.158 &NLS1    &    &  -21.4&  $ 1.45\pm 0.14$& $ 1.8\times 0.5 $  &0.6 & -12.7 & 6.7 \\
3                  & \object {SDSS J134206.57+050523.8} & 0.266 &NLS1    &    &  -23.4&  $ 3.80\pm 0.14$& $ 1.6\times --   $ &0.7 & -12.1 & 8.0\\
4                  & \object {SDSS J150407.51-024816.5} & 0.217 &LINER   &Cl  &  -22.9&  $36.60\pm 0.1$  & $ 2.1\times 1.9 $ &2.0 & -10.9 & 7.5$^{\mathrm{f}}$\\
5                  & \object {SDSS J010649.39+010322.4} & 0.253 &LINER   &Cl  &  -21.4&  $ 1.84\pm 0.15$& $ 6.5\times 4.7 $  &1.5 & -11.6 & 8.6$^{\mathrm{f}}$ \\
6                  & \object {SDSS J074738.38+245637.3} & 0.130 &Sy1     &Comp&  -22.2&  $ 2.47\pm 0.15$& $ 1.6\times --   $ &0.5 & -12.5 & 7.4\\
7                  & \object {SDSS J030639.57+000343.1} & 0.107 &NLS1    &Int &  -21.8&  $ 3.89\pm 0.15$& $ 0.8\times --   $ &0.8 & -12.0 & 7.3\\
8                  & \object {SDSS J081026.07+234156.1} & 0.133 &NLS1    &  &  -21.0&  $ 1.39\pm 0.14$& $ --$ &0.8 & -12.2 & 6.8\\
9$^{\mathrm{e}}$   & \object {SDSS J162332.27+284128.7} & 0.377 &BL Lac  &    &  -22.5&  $ 2.79\pm 0.15$& $ --              $&1.4 & -12.5 & -- \\
10               & \object {SDSS J080322.48+433307.1} & 0.276 &Sy1     &    &  -22.4&  $ 5.25\pm 0.14$& $ 2.3\times 1.0 $  &1.4 & -12.1 & 7.9\\
11$^{\mathrm{e}}$& \object {SDSS J080644.41+484149.2} & 0.370 &Sy1     &    &  -23.8&  $43.31\pm 0.14$& $ 2.7\times 1.9 $  &1.9 & -11.9 & 9.4\\
12               & \object {SDSS J092710.60+532731.6} & 0.201 &passive &Cl  &  -21.1&  $ 3.58\pm 0.22$& $ 2.6\times --   $ &1.8 & -11.8 &  --\\
13& \object {SDSS J134420.87+663717.6} & 0.128 &Sy1    &Int &  -22.9&  $ 2.70\pm 0.40^{\mathrm{g}}$& $67.2\times53.4 $   &0.2 & -11.9 & 7.6 \\
\hline
\end{tabular}
\end{center}
\label{tab:basicParam}
\begin{list}{}{}
\item[$^{\mathrm{a}}$] Spectral classification based on SDSS spectrum \citep[for NLS1 see][]{2006ApJS..166..128Z, 2006AJ....131.1948W}.
\item[$^{\mathrm{b}}$] Int: interacting, Cl: galaxy cluster or group, Comp: companion galaxy as judged from the SDSS images.
\item[$^{\mathrm{c}}$] The FIRST catalog lists the deconvolved major and minor.
Gaussian beam components. Not all deconvolved components are available.
\item[$^{\mathrm{d}}$] Black-hole masses from literature: 990515 from \cite{2004ApJ...614...91W}. The remaining black-hole masses are from \cite{2007ApJ...667..131G}, unless otherwise noted.
\item[$^{\mathrm{e}}$] Source is jet dominated.
\item[$^{\mathrm{f}}$] Using $M_\bullet-\sigma$ relation and [SII] line width as tracer of the bulge stellar velocity dispersion \citep[cf.,][]{2007ApJ...667L..33K}.
\item[$^{\mathrm{g}}$] The radio data originate from the NVSS 20~cm survey.
\end{list}
\end{table*}

Table \ref{tab:basicParam} lists basic source parameters: ID\footnote{For brevity, we will refer to the ID instead of the SDSS name throughout the paper.} (1), SDSS name (2), redshift (3), spectroscopic class (4), local environment (5), extinction-corrected SDSS absolute $i$-band psf magnitude (6), FIRST 20~cm peak flux density in mJy (7), FIRST beam size in sqarcsec (8), logarithm of radio loudness (9), logarithmic units of ROSAT 0.1-2.4~keV flux in erg~s$^{-1}$ (10), and logarithmic units of the black hole mass in $M_\odot$ (11). 
The absolute $i$-band magnitude has been calculated, assuming a power-law\footnote{Throughout the paper we use $F_\nu\propto \nu^{-\alpha}$.} at visible wavelength with index $\alpha_\mathrm{vis}=0.5$ \citep[e.g.,][]{2003AJ....126.2209A} and a $K$-correction of the form $F_i=F_o(1+z)^{\alpha-1}$ \citep{1997iagn.book.....P}.
The radio loudness is defined as $R^*=f_\mathrm{5~GHz}/f_\mathrm{2500\AA}$  \citep[e.g.,][]{1989AJ.....98.1195K}. In the radio domain we use a power-law index $\alpha_r=0.7$ \citep[e.g.,][]{2001AJ....121..128H} in order to estimate the rest-frame 6~cm flux density. The 2500\AA~flux density was estimated from the SDSS $g_\mathrm{PSF}$ magnitude, using the above $\alpha_\mathrm{vis}$. 
Objects clearly above $\log R^*>1$ are usually described as radio loud (RL). Very powerful ones can even reach $R^*>3$ \citep[cf.,][]{2002AJ....124.2364I}. 
X-ray fluxes have been computed by converting the X-ray counts to flux using PIMMS\footnote{Portable, Interactive Multi-Mission Simulator, Version 3.8a2} and  assuming an absorbed power-law with an energy index $\alpha_X=1.5$ \citep[cf.,][]{2003AJ....126.2209A}\footnote{For the energy index we use the form $N(E)\propto E^{-\Gamma}$, and $\Gamma=\alpha+1$.}. The absorbing column densities are Galactic values and were determined from the \ion{H}{I} map of \cite{1990ARA&A..28..215D}.  
Black holes masses have been taken from the literature (see Tab. \ref{tab:basicParam}) and are based on a combination of the measurements of the width of broad components of Hydrogen recombination lines and the luminosity-radius relation of broad-line AGN \citep[e.g.,][]{2005ApJ...629...61K}. 
For (\object{SDSS J150407.51-024816.5} [4] and \object{SDSS J010649.39+010322.4} [5]), we measured the [SII] line dispersions as tracers of the bulge stellar velocity dispersion from their SDSS spectra and employed the $M_\bullet-\sigma_*$ relation as discussed in \cite{2007ApJ...667L..33K}. However, the values are only intended to give a rough idea of the involved black hole masses. \object{SDSS J162332.27+284128.7} [9] is a BL Lac, \object{SDSS J092710.60+532731.6} [12] is a passive galaxy, and both sources show no spectral feature that can be used for SMBH mass-estimates. 

\begin{figure}[ht]
\begin{center}
\resizebox{\hsize}{!}{\includegraphics{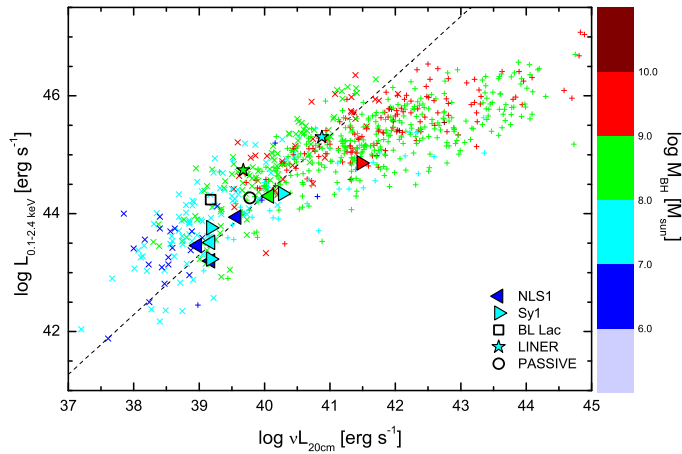}}
\end{center}
\caption{0.1-2.4~keV X-ray vs. 20~cm FIRST peak luminosity. Symbols of our sources are indicated in the legend and correspond to spectroscopic types. Colors are coded according to the black-hole mass. Open symbols have no black-hole mass available. 
For comparison, we present SDSS/FIRST/ROSAT based broad-line AGN (BLAGN) from \citet[$\times$ correspond to RQ and $+$ to RL BLAGN]{2008ApJ...688..826L}. The dashed line is the linear correlation for RQ QSO of \cite{2000A&A...356..445B}. 
}
\label{fig:xr_corr}
\end{figure}

Figure \ref{fig:xr_corr} shows the radio/X-ray correlation for our sample. For comparison we also show the SDSS/ROSAT/FIRST based broad line AGN (BLAGN) sample of \citet{2008ApJ...688..826L}, together with the fit for RQ QSOs from \citep{2000A&A...356..445B}. The BLAGN sample is split into a RQ ($\times$) and a RL ($+$) sub sample. The radio loud objects harbor more massive black holes than the RQ ones, though the dependence on black-hole mass appears to be most intense at lower radio powers \citep[cf., Fig. 4 in][]{2008ApJ...688..826L}.
The RQ correlation has a slope of unity, indicating that the core radio emission is a direct tracer of the nuclear activity. 
Our targets follow the RQ relation closely, with the exception of target 11, this source, however, has a strong, extended jet component. At high radio powers, the RL BLAGN significantly deviate from the RQ relation, indicating that the jet component becomes dominant. If the X-ray emission still originates from (or close to) the accretion disk, while the radio emission is coming from the jet, then a deviation from the RQ correlation is expected.



\section{MERLIN and EVN observations}
\label{sec:obs}
\begin{table}[ht!]
	\caption{Journal of 18~cm observations}
\label{tab:observations}
\centering
\begin{tabular}{ccccc}
\hline
\hline
ID & Date                 & Array       & Exp. time & Phasecal$^{\mathrm{b}}$\\
   & [DD.MM.YY]           &             & [h]            & \\
(1)& (2)                  & (3)         & (4)       & (5)\\
\hline
1  & 04.01.07             & MERLIN       & 1.28 & 1535+004\\
2  & 04.01.07             & MERLIN       & 1.29 & 1502+036\\
3  & 04.01.07             & MERLIN       & 2.77 & 1329+063\\
4  & 04.01.07             & MERLIN       & 1.05 & 1514+004\\
4  & 26.10.08             & EVN          & 3.84 & J1510-0543\\
4$^{\mathrm{a}}$  & 31.10.08& EVN        & 3.85 & J1510-0543\\
5  & 22.12.06, 06.01.07   & MERLIN       & 5.01 & 0106+013\\
6  & 22.12.06             & MERLIN       & 1.93 & 0743+277\\
7  & 22.12.06             & MERLIN       & 1.53 & 0258+011\\
8  & 22.12.06             & MERLIN       & 1.52 & 0802+212\\
9  & 04.\& 05.02.07       & MERLIN       & 13.21& 0552+398\\
10 & 26.\& 27.06.05       & MERLIN       & 6.62 & 0803+452\\
10 & 07.06.07             & EVN          & 2.2  & J0801+4401\\
11 & 28.05.05             & MERLIN       & 9.22 & 0804+499\\
11 & 07.06.07             & EVN          & 2.24 & J0808+4950\\
12 & 27.\& 28.05.05       & MERLIN       & 10.5 & 0929+533\\
12 & 07.06.07             & EVN          & 2.24 & J0932+5306\\
13 & 26.05.05             & MERLIN       & 7.37 & 1337+0637\\
\hline
\end{tabular}
\begin{list}{}{}
\item[$^{\mathrm{a}}$] 6~cm observation.
\item[$^{\mathrm{b}}$] MERLIN phase calibrators names are in IAU B1950 format. EVN phase calibrators names are given in IAU J2000 format.
\end{list}
\end{table}

We acquired 18~cm continuum snapshots of the 13 X-ray bright AGN introduced in Sect. \ref{sec:sample} using MERLIN. 
For four objects we also obtained 18~cm Western EVN (using the stations 
Effelsberg, Medicina, Westerbork, Onsala [85 ft.], Torun, Noto, and Jodrell Bank [Lovell]) observations at higher angular resolution. Target 4 was, furthermore, studied with the same Western EVN at 6~cm.
All observations were carried out in phase-referencing mode, i.e. the target scans were alternated with scans of a nearby phase reference. 
The observations were distributed over several hour angles to ensure a more homogeneous uv coverage. 
Table \ref{tab:observations} gives an overview of the individual observing runs, including the date of observations (2), radio observatory (3), exposure time (4), and the associated phase calibrators (5).
For the MERLIN data, we checked the suitability of the pipeline results (amplitude-, phase calibration, flagging of the phase calibrator) and applied further flagging of the target amplitudes and phases before imaging with the data reduction package AIPS\footnote{http://www.aips.nrao.edu/index.shtml} \citep{1985daa..conf..195W}. 
For the EVN data, we carried out flagging, amplitude/phase calibration and fringe fitting with AIPS before imaging, according to standard reduction recipes\footnote{http://www.evlbi.org/user\_guide/guide/userguide.html}.

\section{Results and discussion}
\label{sec:results}
\begin{figure*}[ht]
\centering
\includegraphics[width=17cm]{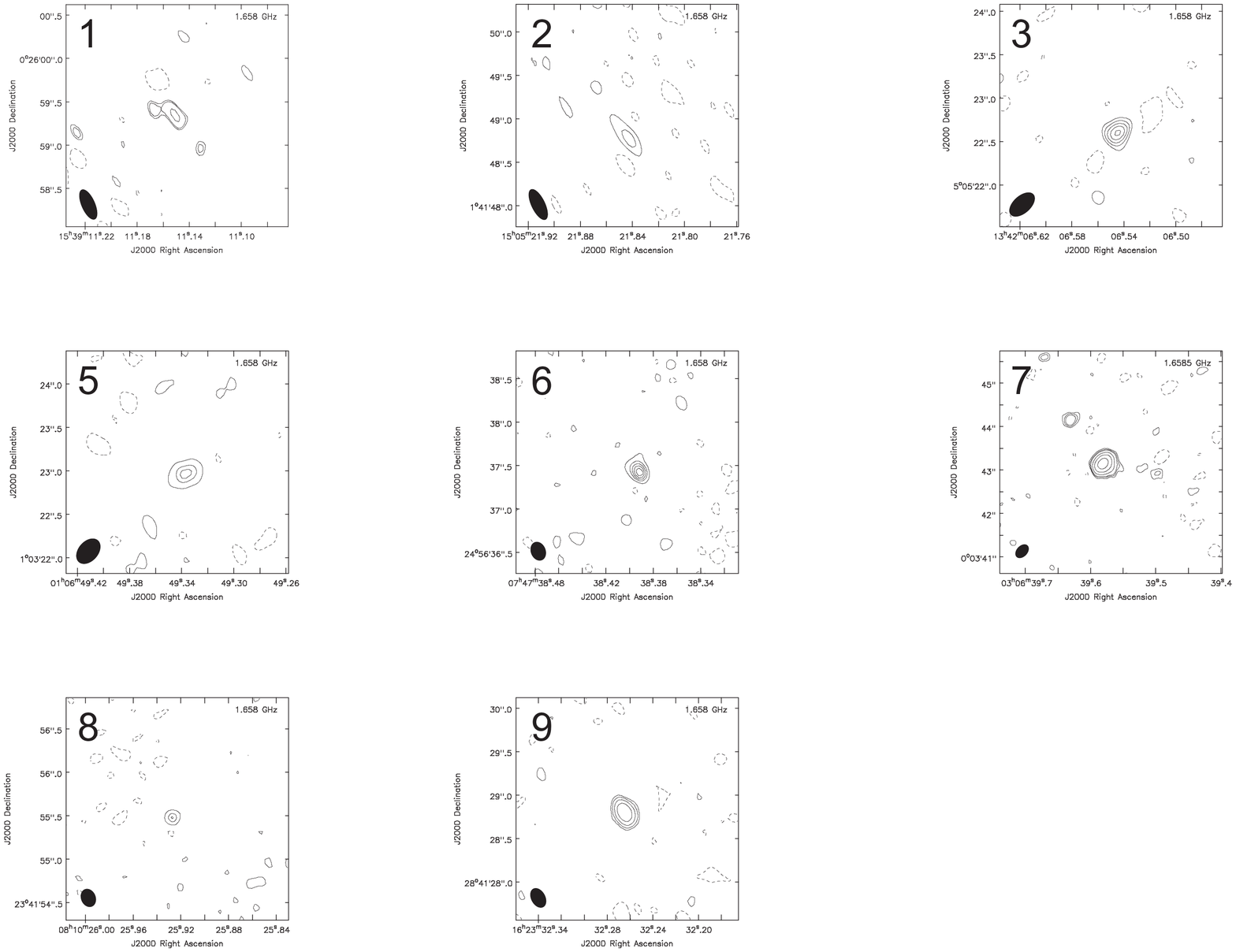}
\caption{Maps of the MERLIN 18~cm reconstructed images of the detected sources (detection threshold $3\sigma$). Contours refer to -2 (dashed), 2, 3, 4, 5, and 6 times the $\sigma$ in the image. Except for sources 7 and 9, for which the contours refer to -2, 2, 3, 4, 8, 16, 32, ... times the $\sigma$. All $\sigma$ values are given in Tab. \ref{tab:derived}. The beams are indicated in the bottom left corner of the maps. North is up, East is left. Sources with EVN follow-up observations are shown in Fig. \ref{fig:maps_evn}.
\label{fig:maps_merlin}}
\end{figure*}

\begin{figure*}[ht]
\centering
\includegraphics[width=17cm]{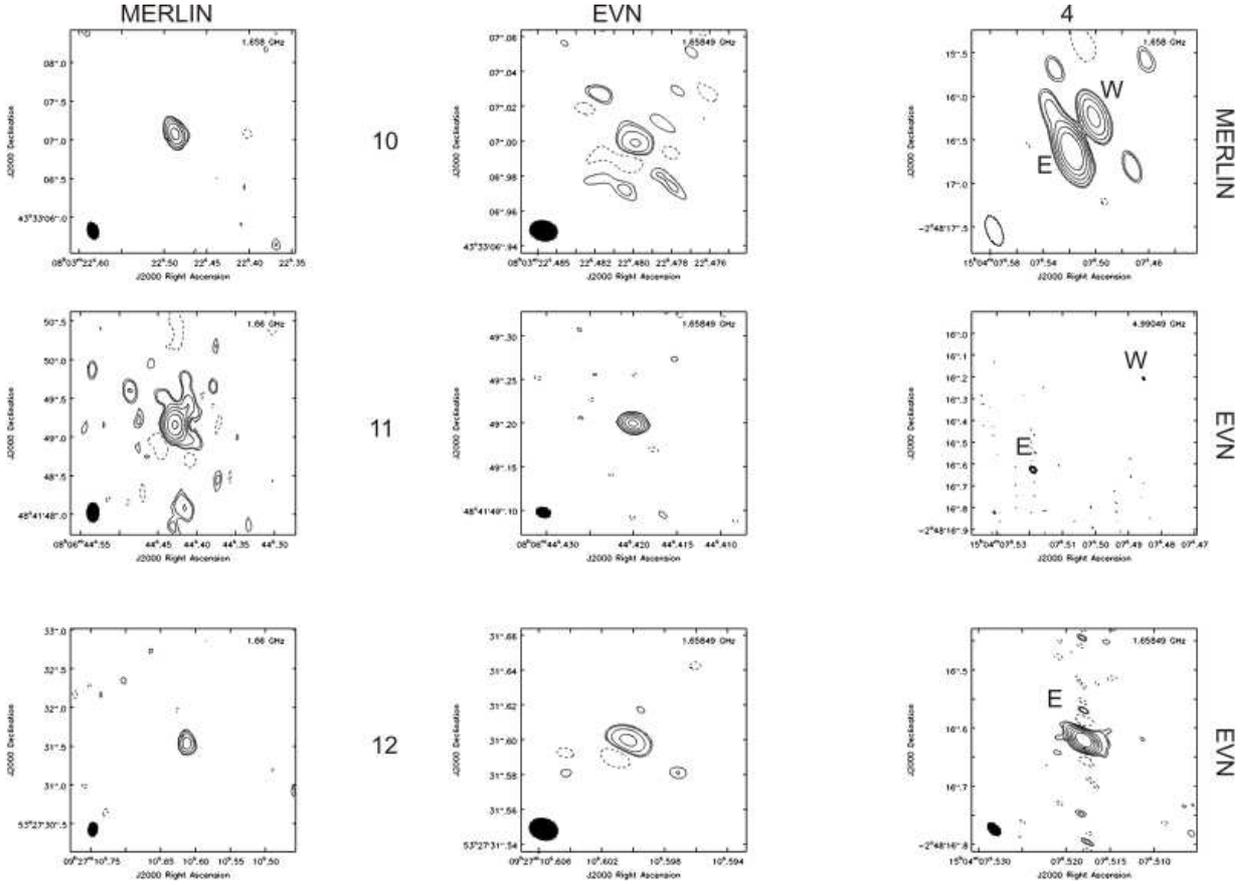}
\caption{
The first two columns present MERLIN and EVN 18~cm reconstructed maps.
The third column shows the MERLIN 18~cm map of source 4 in the first row. The second row shows the EVN 6~cm map including the eastern and western components. In the third row we present a close-up of the eastern component in the EVN 18~cm map. The western component was not detected.
We present maps of the cleaned images of the detected sources (detection threshold $3\sigma$). Contours refer to -3 (dashed), 3, 4, 8, 16, ... times the $\sigma$ in the images. The $\sigma$ values are given in Tab. \ref{tab:derived}. The beams are indicated in the bottom left corner of the maps. North is up, East is left.
\label{fig:maps_evn}}
\end{figure*}

Cleaning and self calibrating the images with Difmap\footnote{ftp://ftp.astro.caltech.edu/pub/difmap/difmap.html} \citep{1997ASPC..125...77S}, employing natural weighting resulted in the Stokes-$I$ maps\footnote{We use the term {\it map}, in the context of the interferometric observations, for the graphical representation of the reconstructed images by contours.} presented in Figs. \ref{fig:maps_merlin} and \ref{fig:maps_evn}. The restoring beam sizes and orientation are indicated in the lower left corner of the maps. Typical beam sizes are $290\times 170$ mas for MERLIN and $19\times 12$ mas for EVN, corresponding to linear scales of about $950\times 560$ pc and $63\times 40$ pc, respectively. The contours represent multiples 
of the standard deviation, $\sigma$, of off-source regions in 
the final reconstructed images, as listed in Table \ref{tab:derived}.
\footnote{The measured $\sigma$ values are consistent with the noise levels, as derived theoretically \citep[cf.,][]{1999ASPC..180..171W}, using typical MERLIN system parameters.}
The table furthermore lists source properties like peak flux density, major and minor axes of a Gaussian fit to the central component, and the extent of the central component, measured as $\epsilon=\sqrt{F_\mathrm{int}/F_\mathrm{peak}}$ \citep[cf.,][]{2002AJ....124.2364I}. Sources with $\epsilon>1$ are extended.


All targets, except \object{SDSS J134420.87+663717.6} [13], have been detected with MERLIN and EVN. Noise levels range from $0.07-0.6$~mJy in the MERLIN images and $0.06-1.14$~mJy in the EVN images. As can be deduced from the maps, all detected sources are dominated by a nuclear component. The extent parameter indicates that these nuclear components are marginally resolved ($\epsilon\sim 1.16$).
Note that the boundary between unresolved and resolved source is not sharp, since errors in the estimation of peak and integrated flux densities have to be taken into account. Assuming 10\% errors of the flux measurements (which is the mean error for all our measurements), results in a $\sim 10\%$ error for $\epsilon$. This means that $\epsilon=0.9$ and $\epsilon=1.1$ have the same meaning. 
Targets \object {SDSS J030639.57+000343.1} [7] and \object {SDSS J080644.41+484149.2} [11] show the most extended structures according to their $\epsilon$. 
A special case is target [4], for which we detected a second strong nuclear component (also see App. \ref{sec:notes} for further discussion). This is the only source that was also targeted at 6~cm. Both components were detected at 6~cm, but only the stronger component was detected at 18~cm (Fig. \ref{fig:maps_evn}). Bad weather conditions during the 18~cm run could have partially been responsible for a reduced sensitivity outside the image center.
\begin{sidewaystable*}
\begin{minipage}[t][0mm]{\textwidth}
\caption{MERLIN, EVN measurements, basic source parameters, and derived brightness temperatures, star-formation rates.} 
\label{tab:derived}
\centering
\begin{tabular}{llcccccccccccccc} 
\hline\hline             
Source&                    &1  &2       & 3   &4E   &4W   &5    &6    & 7   &8    &9    &10 &11 & 12& 13\\
\hline
\textsf{MERLIN 18~cm}\\
\cmidrule(r){3-12}\cmidrule(r){13-16}
&&\multicolumn{10}{c}{2006}&\multicolumn{4}{c}{2005}\\

\hline
$F_\mathrm{peak}$ 	& [mJy/beam]   &3.4  &1.10	&3.1	&56.5	&13.8	&0.45	&0.49	&3.00	&0.38	&2.20	&4.14	&37.50	&2.10	&$<0.07$ \\
$\sigma$		& [mJy/beam]   &0.6  &0.19	&0.5	&0.30	&0.30	&0.09	&0.07	&0.06	&0.09	&0.07	&0.09	&0.25	&0.08	&0.07 \\
$\theta_\mathrm{major}$&[arcsec] &0.38 &0.50	&0.34	&0.36	&0.35	&0.32	&0.21	&0.34	&0.14	&0.22	&0.22	&0.25	&0.18	&... \\
$\theta_\mathrm{minor}$&[arcsec] &0.20 &0.18	&0.29	&0.16	&0.16	&0.22	&0.15	&0.30	&0.12	&0.13	&0.13	&0.17	&0.12	&... \\
$\epsilon$             &         &1.19 &1.12   &1.16   &1.19   &1.19  &1.14   &1.16   &1.24   &1.0    &1.14   &1.17   &1.21   &1.16   &... \\
\hline
\textsf{EVN 18~cm (6~cm)}$^{\mathrm{a}}$\\
\cmidrule(r){6-7}\cmidrule(r){13-15}
\multicolumn{5}{c}{}&\multicolumn{2}{c}{2008}& \multicolumn{5}{c}{}&\multicolumn{3}{c}{2007}\\
\hline
$F_\mathrm{peak}$ 	 & [mJy/beam]  &...&	...	&...	&46.8 (40.0)	&$<1.4$ (0.74)	&...	&...	&...	&...	&...	&5.35   &33.60	&5.31	&... \\
$\sigma$	 	 & [mJy/beam]  &...&	...	&...	&1.4 (0.06)     &1.4 (0.06)	&...	&...	&...	&...	&...	&0.10	&0.10	&0.10	&... \\
$\theta_\mathrm{major}$ &[arcsec] &...&	...	&...	&0.024 (0.013)	&... (0.012)	&...	&...	&...	&...	&...	&0.020  &0.017	&0.023	&... \\
$\theta_\mathrm{minor}$ &[arcsec] &...&	...	&...	&0.013 (0.006)	&... (0.006)	&...	&...	&...	&...	&...	&0.013  &0.012	&0.013	&... \\
$\epsilon^{\mathrm{c}}$&         &...& ...       &...     &1.23 (1.21)    &... (1.17)      &...	&...	&...	&...	&...	&1.19   &1.24   &1.18 & ...\\
\hline
~\\
\hline
$F^\mathrm{MERLIN}_\mathrm{peak}/F^\mathrm{FIRST}_\mathrm{peak}$& [\%]&306.3 &75.9  &81.6  &152.9  &37.7 &24.5  &19.8  &77.1  &27.3  &79.6  &78.9  &86.6  &58.7  &...\\
$F^\mathrm{EVN}_\mathrm{peak}/F^\mathrm{MERLIN}_\mathrm{peak}$& [\%]&...    &...    &...    &83.6   &...   &...    &...    &...    &...    &...    &129.2 &89.6  &252.9 &...\\
$\log L_\mathrm{20~cm}$ & [erg s$^{-1}$ Hz$^{-1}$]  &30.4&30.0	&30.9	&31.7	&31.7	&30.5	&30.0	&30.0	&29.8	&31.1	&31.1	&32.3	&30.6	&30.0\\
$\log L_\mathrm{MERLIN}$ & [erg s$^{-1}$ Hz$^{-1}$]&30.8&29.9	&30.8	&31.3	&31.9	&29.9	&29.3	&29.9	&29.2	&31.0	&31.0	&32.2	&30.4	&...   \\
$\log L_\mathrm{EVN}$ & [erg s$^{-1}$ Hz$^{-1}$]   &...	&...	&...	&31.8	&...	&...	&...	&...	&...	&...	&31.1	&32.2	&30.8	&...   \\
$\log L_\mathrm{0.1-2.4~keV}$ & [erg s$^{-1}$]      &43.9	&43.2	&44.3	&45.3	&45.3	&44.7	&43.2	&43.5	&43.4	&44.2	&44.3	&44.8	&44.2	&43.7\\
$\log T_\mathrm{b,20~cm}$ & [K]                     &1.8$^{\mathrm{b}}$	&3.1	&3.1$^{\mathrm{b}}$	&3.9	&3.9	&1.7	&2.9$^{\mathrm{b}}$	&3.6$^{\mathrm{b}}$	&2.3$^{\mathrm{b}}$	&2.5$^{\mathrm{b}}$	&3.3	&3.9	&2.6$^{\mathrm{b}}$	&-0.3  \\
$\log T_\mathrm{b,MERLIN}$ & [K]                   &4.6	&3.9	&4.4	&5.9	&5.3	&3.7	&4.0	&4.3	&4.2	&4.8	&5.1	&5.9	&4.9	&...   \\
$\log T_\mathrm{b,EVN}$ & [K]                       &...	&...	&...	&8.1(8.5)&... (6.9)&...	&...	&...	&...	&...	&7.2	&8.2	&7.1	&...   \\
$\log$ SFR$_\mathrm{20~cm}$ & [$M_\odot$ yr$^{-1}$]  &2.1	&1.7	&2.7	&3.4	&3.4	&2.3	&1.7	&1.7	&1.5	&2.8	&2.8	&4.0	&2.3	&1.7  \\
$\log$ SFR$_\mathrm{MERLIN}$ & [$M_\odot$ yr$^{-1}$]&2.5&1.6	&2.5	&3.6	&3.0	&1.5    &1.0	&1.6	&1.0	&2.7	&2.7	&3.9	&2.1	&...   \\
$\log$ SFR$_\mathrm{EVN}$ & [$M_\odot$ yr$^{-1}$]  &...	&...	&...	&3.5 &...&...	&...	&...	&...	&...	&2.8	&3.9	&2.5	&...   \\
$\log$ SFR$_\mathrm{[OII]}$ & [$M_\odot$ yr$^{-1}$]  &1.4	&0.5	&1.4	&1.9 & 1.9 & 1.7& 0.8	&0.3	&0.4	&...     &0.0	&1.6	&-0.4	&0.9\\
SNR$_\mathrm{MERLIN}$ & [yr$^{-1}$]                & 3.3 & 0.4   & 3.3   & 42.0& 10.6& 0.4 & 0.1 & 0.4 & 0.1 & 5.3 & 5.3 & 83.8& 1.3& ... \\
SNR$_\mathrm{EVN}$ & [yr$^{-1}$]                &...& ...& ...& 33.4& ...& ...& ...& ...& ...& ...& 6.7 & 83.8& 3.3 & ...\\
\hline
\end{tabular}
\begin{list}{}{}
\item[$^{\mathrm{a}}$] Values in brackets correspond to 6~cm EVN measurements.
\item[$^{\mathrm{b}}$] In case one component (major or minor axis) of the source size is missing, the existing one is used. In case both components are missing, the median source size is used, according to $\theta\approx 2F_{20~cm,\mathrm{peak}}^{0.3}$\,arcsec \citep[cf.,][]{1995ApJ...450..559B}.

\end{list}
\vfill
\end{minipage}
\end{sidewaystable*}

	Fig. \ref{fig:sfr_tb} displays the brightness temperature versus the radio-derived star-formation rate (SFR).
	As discussed below (Sect. \ref{sec:importanceofstarformation}), the SFR is proportional to the radio luminosity density. Therefore, the figure also displays
the variation of peak flux density with used array (VLA - MERLIN - EVN; i.e., the probed angular scales).
For most sources, a trend of decreasing peak flux densities of the central marginally resolved point source with increasing angular resolution is evident. This is consistent with the idea of attributing part of the radio emission (up to 50\% from FIRST to MERLIN) to extended jet structures or star-formation. These features are just resolved out by the longer baselines. 
Variability of the nuclear source can also account to some degree for the differences between the multi-epoch/multi-aperture data. As will be discussed in the next Section, the increased peak flux densities for some sources (1, 4, 10, 12) at later epochs clearly point towards variability. For those sources, where the peak flux density at later epochs is decreasing, we cannot clearly distinguish between the two effects. 

\subsection{Extended emission}
\label{sec:extendedemission}
The sources of this sample - except target 11, which shows a compact core plus two distant lobes - are unresolved on FIRST scales. 
We can, therefore, compare the integrated flux density in the MERLIN images, $F_\mathrm{peak}\times\epsilon^2$, with that of the FIRST peak flux densities (Tab. \ref{tab:derived}).\footnote{For a typical radio spectral index ($\alpha=0.7$), the MERLIN 18~cm fluxes densities are only marginally different from the estimated 20~cm ones.}
Four (5, 6, 8, 12) of the 11 MERLIN detections (ignoring the double source 4 for the moment) are clearly fainter on MERLIN than on FIRST scales, indicating resolved extended emission. The resolved fractions reach up to 70\%. Another four sources (2, 3, 9, 10) have integrated MERLIN flux densities similar to the FIRST peak flux density, indicating that most of the flux is still contained in the compact nuclear component. The remaining three sources (1, 7, 11) are clearly brighter on MERLIN scales, indicating for variability. 
In addition to source intrinsic processes, interstellar scintillation in our Galaxy can be responsible for part of the observed variability \citep[e.g.,][]{2011piim.book.....D}. 
This is particularly important in the 3-8~GHz regime and due to changing sight lines through the interstellar medium to the target during the motion of the Earth around the sun. 
It has been observed that the variability amplitude of compact (on VLBI scales) cores of QSOs related to this phenomenon is typically up to 25\% \citep[e.g.,][]{1992A&A...258..279Q}.
We, therefore, checked the 60~$\mu$m and 100~$\mu$m IRAS maps around the positions of our sample objects for Galactic cirrus emission \citep[cf.,][]{1987A&A...183..335J} and the maps for local (within a few parsecs) turbulent warm interstellar clouds \citep{2008ApJ...675..413L}.
We find that scintillation potentially contributes to the variability of our sources. In terms of differential cloud velocity \citep{2008ApJ...675..413L}, sources 10 and 12 are the least, and source 1 is most stronly affected by scintillation.  
Without any further correction for this effect, the measured core fluxes of our targets are constant on that level, and excess variation is most likely due to source intrinsic properties (i.e., extended emission or variability of the central engine).
Nevertheless, our basic conclusions regarding the radio loudness and the location in the fundamental plane, as discussed in the following sections, are not hampered by this effect. 
Further (pseudo-simultaneous) multiwavelength observations are required to resolve the ambiguity between variability due to jet/AGN or supernovae or interstellar scintillation contributions.

\subsection{Importance of star formation}
\label{sec:importanceofstarformation}
\subsubsection{Brightness temperatures}
The brightness temperature, $T_B$, helps distinguishing between emission mechanisms.
The angular size of the restoring beam can be used to determine $T_B$, corresponding to the unresolved nuclear radio flux:
\begin{equation}
	T_B=\frac{c^2I_\nu}{2k\nu^2}=1.8\times 10^9(1+z)\left(\frac{S_\nu}{\mathrm{mJy}}\right) \left(\frac{\nu}{\mathrm{GHz}}\right)^{-2} \left(\frac{\theta_\mathrm{major}\theta_\mathrm{minor}}{\mathrm{mas}^2}\right)^{-1}\mathrm{K}
\end{equation}
where $c$ is the speed of light, $\nu$ the frequency, $k$ the Boltzmann constant, and $I_\nu$ the specific monochromatic intensity, which is the specific flux density divided by the solid angle $\Omega$, the extent of the radio emitter. The solid angle is approximately $\Omega\approx\theta_\mathrm{major}\times\theta_\mathrm{minor}$ of the unresolved source component. The factor $(1+z)$ accounts for the redshift of the source. 
The typical temperature of an HII region is of the order of $10^4$~K.
Brightness temperatures well above $10^4$~K \citep[typical for star-forming galaxies, e.g.,][]{1982ApJ...252..102C}, are unlikely due to star formation \citep[cf., discussion in][]{2007A&A...464..553K}. The effective temperature of radiation for classical accretion disks is in the range $10^5-10^7$~K \citep[cf.,][]{1998tbha.conf..148N}.
Brightness temperatures in excess of $10^7$~K are related to nonthermal processes, at very high levels ($>10^9$~K) supposedly to relativistically beamed jet emission \citep[e.g.,][]{1994ApJ...426...51R}.
Table \ref{tab:derived} lists $T_B$ for the individual measurements (FIRST, MERLIN, EVN). The brightness temperatures for given angular size vary between $10^{2-4}$\,K for FIRST, $10^{4-6}$\,K for MERLIN, and $10^{7-8}$\,K for EVN; clearly increasing with increasing angular resolution. Brightness temperatures of the unresolved VLBA cores of RQQs are about $10^{8-9}$\,K \citep{2005ApJ...621..123U}, whereas luminous quasars can exhibit brightness temperatures of up to $10^{12}$\,K \citep[e.g.,][]{1969ApJ...155L..71K}. 
At least for the EVN data the values are too high to be dominated by star formation. 

\subsubsection{Star-formation indicators}
\label{sec:starformation}

A large body of work on star-forming galaxies shows correlations between X-ray/FIR \citep[e.g.,][]{1992ApJ...388...82D}, and radio/FIR \citep[e.g.,][]{1991ApJ...376...95C}, as well as calibrations of supernova and star-formation rates \cite[e.g.,][]{1992ARA&A..30..575C, 2001AJ....121..128H, 2003ApJ...599..971H,2003ApJ...586..794B}, which allow us to test the hypothesis that the nuclear emission is dominated by star formation.

First, we can estimate star-formation rates and FIR luminosities from our radio measurements.
We will use the 18~cm flux densities directly, since, for reasonably flat spectra of radio emission, the differences between 18~cm and 20~cm flux densities are negligible.
Star-formation rate is, according to \cite{2003ApJ...586..794B}, related to the 20~cm luminosity density by
\begin{equation}
	\left(\frac{\mathrm{SFR}}{M_\odot\mathrm{yr}^{-1}}\right) = 5.52\times 10^{-15}\left(\frac{L_\mathrm{20cm}}{\mathrm{erg\,s}^{-1}\mathrm{\,Hz}^{-1}}\right)
\label{eq:radioSFR}
\end{equation}
The estimated SFRs for the FIRST, MERLIN, and EVN measurements are listed in Tab. \ref{tab:derived}. For about half of the sources the SFRs are $>100$ and reach up to $10^4\,M_\odot\,$yr$^{-1}$. \cite{2003ApJ...599..971H} find SFRs $\lesssim 100\,M_\odot\,$yr$^{-1}$ for most of their SDSS-based sample of inactive galaxies.
There are extreme examples of starbursts/interacting galaxies with SFRs of up to $1000\,M_\odot\,$yr$^{-1}$ \citep[cf.,][]{1998ARA&A..36..189K}. These objects, however, are ultraluminous infrared galaxies \citep[ULIGs, e.g.,][]{1996ARA&A..34..749S}. 
We can also estimate the infrared (IR) flux in case that all radio emission is due to star formation, i.e. re-radiation of hot dust.
We use the radio-SFR in Eq. (4) of \cite{1998ARA&A..36..189K}. For most of our objects the estimated IR luminosity falls in the regime of ULIGs. As such, they should have been detected by IRAS.
\footnote{Take, for example, the ULIG 3C~48, which has a similar redshift and infrared luminosity ($z=0.3695$, $L_\mathrm{IR}\approx 5\times 10^{12}L_\odot$) and has been clearly detected with IRAS \citep{1985ApJ...295L..27N}.}

Supernovae are the end product of massive star formation. There appear to be exceptionally bright type-2 radio supernovae (RSNe). Following \citep{2001AJ....121..128H}, we compare our core flux densities with that of one of the brightest supernovae, SN1986J in NGC~891 \citep{1990ApJ...364..611W} shifted to the average redshift of our sample ($z\sim 0.2$). The 20~cm and 6~cm light curves peak at about 130~mJy. Our sample is about 1000 times more distant than NGC~891. In order to contribute significantly to the unresolved core emission, of the order of 10-100 RSNe have to be involved. This is quite high, especially when considering the small volume sampled by the EVN observations \citep[cf.,][]{1992MNRAS.255..713T,1994MNRAS.266..455M}. Furthermore, assuming that all of the radio flux has a non-thermal origin, we can estimate the type-II supernova rate (SNR) according to \cite{1990ApJ...357...97C}:
\begin{equation}
\left(\frac{\mathrm{SNR}}{\mathrm{yr}^{-1}}\right)\approx 7.7\times 10^{-31}\left(\frac{\nu}{\mathrm{GHz}}\right)^\alpha \left(\frac{L_\mathrm{NT}}{\mathrm{erg\,s}^{-1}\mathrm{Hz}^{-1}}\right),
\end{equation}
where $\alpha\sim -0.8$ is the nonthermal radio spectral index and $L_\mathrm{NT}$ is the nonthermal component of the radio emission. Tab. \ref{tab:derived} lists the estimated SNRs, which range from 0.1-83 yr$^{-1}$. Starburst galaxies show SNRs of the order of $\lesssim 10$ yr$^{-1}$ \citep[e.g.,][]{2003A&A...401..519M}, whereas Seyferts, with some exceptions, show SNRs of the order of $\lesssim 1$ yr$^{-1}$ \citep[e.g.,][]{2001AJ....121..128H,2006AJ....131..701L}. 

Furthermore, we can use the the [OII]$\lambda 3728\AA$ luminosity, measured from the SDSS spectra, to estimate SFRs \citep[e.g.,][]{2003ApJ...599..971H,1998ARA&A..36..189K}.
It has been demonstrated that [OII] can also be used as a star-formation tracer in AGN \citep{2005ApJ...629..680H}, and we will show, that, consistently, star formation is not significantly contaminating the nuclear emission. 
In high ionization type-1 AGN, flux ratios of [OII]/[OIII] typically range from 0.1-0.3 \citep[][and references therein]{2005ApJ...629..680H}. 
Excess [OII] emission can then be attributed to star formation \citep[cf.,][]{2008MNRAS.390..218V}. 
In our sample the median [OII]/[OIII]$=0.53$, ranging from 0.13 to 4.9. Except for the two LINER galaxies [4, 5], which show the highest ratio in excess of a factor of about 15 over the typical type-1 AGN value, the ratios are quite similar to type-1 ionization. 
We follow the prescription of \cite{2005ApJ...629..680H} in estimating [OII]-based SFRs.
At this point we neglect excitation of [OII] via shocks. These might be important for LINERs \citep{1980A&A....87..152H} and in such cases, we overpredict the [OII]-based SFRs. 
With the \cite{1998ARA&A..36..189K} calibration, we then find a median SFR of about $7.3\,M_\odot$~yr$^{-1}$, ranging from 0.4 to $71\,M_\odot$~yr$^{-1}$. 
This demonstrates that star formation contributes to the overall emission. However, inverting Eq. \ref{eq:radioSFR} to estimate radio luminosities based on the [OII]-derived SFRs and comparing with Tab. \ref{tab:derived}, especially with those SFRs derived from our MERLIN and EVN data, shows that star formation cannot explain the compact radio emission of our sources. The [OII]-based radio luminosities are at least an order of magnitude below those of the observed radio luminosities. 
There appear to be two exceptions, sources [5] and [6], for which [OII] and radio SFRs are of the same order of magnitude and are comparably small.
Target [5] is one of the two LINER galaxies in our sample. The fact that the SFRs are of the same order of magnitude might be misleading, since part of the [OII] emission is likely to be excited by shocks as mentioned above.
Note that the 3\arcsec diameter SDSS fibers probe galactic scales of the order of 10~kpc at the average redshift of our sample, whereas MERLIN probes ten times smaller scales. In that sense, the SDSS fibers much better couple to the FIRST beam size of about 5\arcsec and probably much of the star formation is extended throughout the host galaxies. A direct comparison of the MERLIN and EVN based with the [OII] based SFRs has to taken with caution and should only be regarded as a coarse indicator for the contribution of star formation to the radio emission.

\begin{figure}[ht!]
\begin{center}
\resizebox{\hsize}{!}{\includegraphics{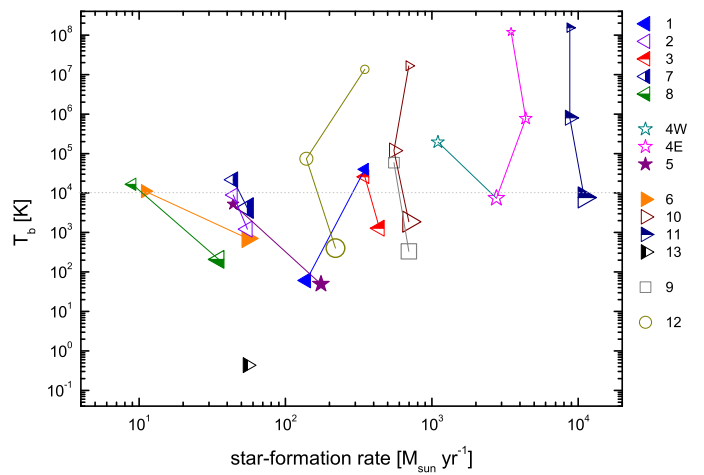}}
\end{center}
\caption{Brightness temperature vs. star-formation rate, both estimated from radio flux densities. The symbol size corresponds to the source size of the individual observation. FIRST, MERLIN, and EVN measurements of each source are connected by a line.
General symbol shapes as in Fig.
\ref{fig:xr_corr}.}
\label{fig:sfr_tb}
\end{figure}

Fig. \ref{fig:sfr_tb} summarizes the situation based on brightness temperature and radio SFR. With higher angular resolution (MERLIN) part of the emission is resolved out.
The exceptions mentioned at the end of the previous paragraph consistently are on the left side of the diagram. 
With still increasing angular resolution (EVN), however, the nuclear emission is rather unresolved, the SFR remains constant, and the brightness temperature increases. 
The EVN cores are, therefore, likely due to nonstellar nuclear processes, being consistent with the AGN nature of the visible spectra.   

\subsection{The radio/X-ray connection}
In the previous section we have seen that for most of the sample objects the compact (i.e., unresolved) radio emission is of non-stellar origin. 
We can combine our high angular resolution 18~cm observations with archival soft X-ray data to bring our sources in context with the radio/X-ray correlation.
We will show that the sources of our sample follow the general trends of the X-ray/radio correlation and that they are consistent with the fundamental plane relation of \cite{2003MNRAS.345.1057M}, which rather traces the accretion flow than the connection with the (radio) jet.
Since the X-ray and radio data are not from the same epoch, we cannot exclude that variability has affected the fluxes in each of the components. Here, we assume that the measured fluxes are representative for the average flux levels of the presented AGN.
At radio frequencies (Fig. \ref{fig:sfr_tb}), all but two sources present decreasing flux-density values on MERLIN scales, while on EVN scales mostly the level of the FIRST flux density is reached.
Considering the SDSS/ROSAT/FIRST BLAGN sample of \cite{2008ApJ...688..826L}, the large number of objects provides statistical trends for the population. 
As will be discussed in the following paragraphs, the objects presented in this work are following these general trends, indicating that variability does not have a significant impact.
Concerning the X-ray domain, three objects (4, 5, and 12) are located close to/at the centers of clusters of galaxies. 
Therefore, a significant fraction of the X-rays originates from the hot intracluster gas and will bias our results (see App. \ref{sec:notes}).

\begin{figure}[ht]
\centering
\resizebox{\hsize}{!}{\includegraphics{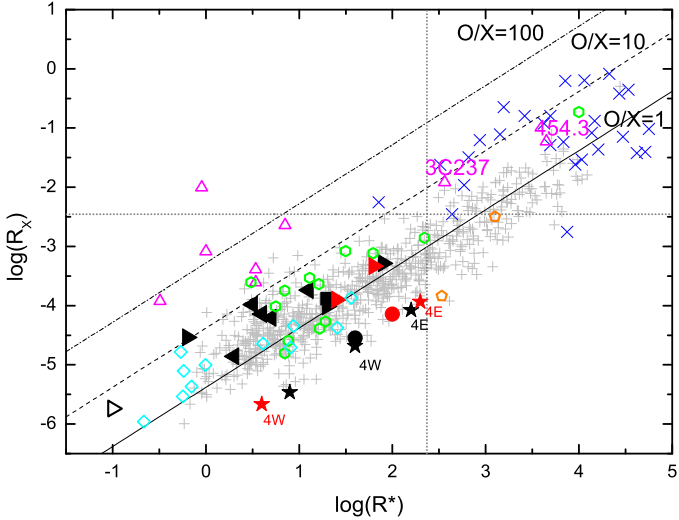}}
\caption{Radio/X-ray vs. radio/optical flux ratio. Diagonal lines indicate optical/X-ray ratios of 1, 10, and 100. Symbol shapes as in Fig. \ref{fig:xr_corr}. Black symbols are based on our MERLIN measurements. The black open right triangle corresponds to the radio upper-limit based values of source 13. Red symbols are based on our EVN measurements. The open magenta triangles are data from \cite{2005ApJ...621..123U}. Green hexagons are low luminosity Seyfert 1,2 from \cite{panessa_x-ray_2007}, blue crosses radio galaxies from \cite{2006A&A...447...97B}. 
Cyan diamonds are soft X-ray selected NLS1s from \cite{2004AJ....127..156G}.
Light grey plus signs are BLAGN from \cite{2008ApJ...688..826L}. Japanese VLBI observations of two of them are represented by orange pentagons \citep{2007PASJ...59..703D}. Note that only our data as well as those of \citeauthor{2005ApJ...621..123U} and \citeauthor{2007PASJ...59..703D} are based on high angular resolution radio observations. The dotted horizontal and vertical lines mark the refined RL/RQ demarcation from \cite{panessa_x-ray_2007}.}
\label{fig:RXRO}
\end{figure}

\subsubsection{Radio loudness}
Radio loudness and the nature of its apparent dichotomy (RL and RQ AGN) are still a matter of debate. Selection effects seem to play an important role. 
However, when conducting deep radio observations, AGN appear to show a rather continuous distribution of the radio loudness \citep[e.g.,][]{2003MNRAS.346..447C}.
Nevertheless, work on large SDSS-based samples still recovers a bimodality \citep{2002AJ....124.2364I}.
RL AGN are associated with the largest black-hole masses and early-type host galaxies, while late-type galaxies almost exclusively host RQ AGN \citep[e.g.,][]{2010MNRAS.tmp.1111H}. 
On the other hand, using high angular resolution in the radio (e.g., VLA) and the visible (e.g., HST), even classical low redshift RQ AGN and low luminosity AGN (despite their low intrinsic power) turn out to have RL cores \citep{2001ApJ...555..650H}. 
This emphasizes the importance of angular resolution and sensitivity for a thorough classification.

\cite{2003ApJ...583..145T} introduced a new measurement of the radio loudness. They compute the ratio of 5~GHz over hard X-ray flux, $RX=\nu S_\nu(\mathrm{5~GHz})/S(\mathrm{2-10~keV})$. The use of X-ray flux should overcome extinction effects that are important at visible wavelengths. The unresolved nuclear X-ray source is also more reliably connected to the AGN than the visible emission, which could be contaminated by host galaxy light. 
Figure \ref{fig:RXRO} displays the ratio of radio over X-ray luminosity versus classical radio loudness. Also indicated are lines of constant optical over X-ray luminosity. The dotted horizontal and vertical lines show the refined RQ/RL separation by \cite{panessa_x-ray_2007}. 
Our data points are shown as filled black (MERLIN) and red (EVN) symbols. The shapes of the symbols are as in Fig. \ref{fig:xr_corr}. 
Our data points populate the RQ region between $1\leq O/X\leq 10$. The two LINERs (4 and 5, filled stars) and the passive galaxy (12, filled circle) show a relative increased X-ray contribution due to X-rays from the harboring galaxy clusters. 
The large ROSAT beam does not allow us at this point to decompose the X-ray emission into AGN and cluster emission.
As discussed above, in 11 of 13 cases, the MERLIN observations resolve out 15-80\% (with an average of about 41\%) of the FIRST detected emission (assuming no contribution by variability). 
This means that the AGN become more radio quiet, and this effects moves the data points along $O/X=const$ lines; e.g. a 50\% resolved fraction of the MERLIN data would move the corresponding FIRST data points by $-0.3$dex along $\log R_X$ and $\log R^*$.

For comparison, we present data from various related studies. 
We actually show the 0.1-2.4~keV instead of the 2-10~keV fluxes. As noted by \cite{2005ApJ...621..123U}, applying the commonly used X-ray photon index of $\Gamma=2$ in computing the 2-10~keV flux results in an average reduction of the X-ray flux by a factor of 2. 
We, therefore, multiplied the 2-10~keV fluxes of the comparison objects by that factor, missing knowledge on the X-ray spectra justifies such a procedure.\footnote{Here, a certain amount of caution is advisable, because it is currently unclear, whether soft (0.1-2~keV) and hard X-rays (2-10~keV) have the same origin for the various classes of objects presented in Figs. \ref{fig:RXRO} and \ref{fig:merloni}. Furthermore, soft 0.1-2.4~keV X-rays are more prone to extinction than the harder 2-10~keV X-rays.}
Small, grey symbols show SDSS/ROSAT/FIRST-based BLAGN ($<z>=0.67\pm0.49$ for $0.02<z<2.17$) from \cite{2008ApJ...688..826L}. Black hole masses range from $10^{6-8}M_\odot$ ($<M_\bullet>=3\times10^8M_\odot$).
The data span four decades of radio loudness and closely follow the $O/X=1$ line. 
According to the revised RQ/RL separation by \cite{2007A&A...467..519P} about 12\% of their sample can be classified as RL. In contrast, using the classical definition, 69\% of their sample is RL.
The nearby ($z< 0.015$) low luminosity AGN (type-1, type-2 and LINER/HII) from \cite{panessa_x-ray_2007} are shown as green hexagons. Black hole masses are in the range $10^{5-8}M_\odot$ with a peak just below $10^8M_\odot$ \citep{2006A&A...455..173P}. 
Open cyan diamonds are soft X-ray selected NLS1 galaxies (with $0.017<z<0.27$) from \cite{2004AJ....127..156G}. The RQ NLS1s have likely rather small black-hole masses of around $\lesssim 10^6M_\odot$\footnote{Based on the width of their broad H$\beta$ lines and the broad-line-region radius luminosity relation from \cite{2009ApJ...697..160B}.}.
Blue crosses are core radio galaxies and low luminosity radio galaxies (with $0.003<z<0.01$) from the study of \cite{2006A&A...447...97B}. 
These objects are clearly radio loud and have a mean black hole mass of $\sim4\times10^8M_\odot$.
Among high angular resolution studies, we show QSOs studied with VLBA (on mas scales) by \citet[open, magenta, upright triangles]{2005ApJ...621..123U}. The RQ QSOs \citep[based on the samples of][]{1996AJ....111.1431B,1997AJ....113..144B} are mostly unresolved on VLBA scales, exhibit brightness temperatures of about $10^9$K, and have typical black hole masses of around $10^9M_\odot$. 
The RQ QSOs clearly populate a different region ($O/X\approx 100$) than the other RQ AGN presented in the figure. 
The RL QSOs are similar in their radio loudness characteristics to those from the other studies. 
Orange pentagons are JVN observations of two NLS1s by  \cite{2007PASJ...59..703D}. They have black hole masses of about $10^7M_\odot$ and are also part of the \cite{2008ApJ...688..826L} sample. The JVN data are spatially unresolved on mas scales and brightness temperatures in excess of $10^7$K.

In Fig. \ref{fig:RXRO}, most of the data points lie between $1\lesssim O/X\lesssim 10$, especially those of the BLAGN from \cite{2008ApJ...688..826L} scatter around the $O/X=1$ line. 
In the RQ domain, \cite{2005ApJ...621..123U} find slight indications for increasing $O/X$ for decreasing $R^*$.
The authors discuss that this effect could be related to an increased X-ray contribution by inverse Compton up-scattered photons from the base of the jet for RL objects.
For RL objects, especially radio galaxies, $O/X$ can again reach values $\ga 10$.
Interestingly, the RQ objects with the highest $R_X$ are broad absorption line QSOs, which are known to be X-ray faint and it is still being a matter of discussion, whether the X-ray faintness is due to extinction \citep[e.g.,][]{2009ApJ...702..911M,2008MNRAS.390.1229B}.
In case extinction is important, $R^*$ in these objects could be biased and corrected values could actually be located in the RL domain.
One has to emphasize the importance of thoroughly studying the nuclear radio to X-ray emission in order to derive the intrinsic nuclear continuum \citep[cf.,][]{2001ApJ...555..650H}.
Angular resolution varies with telescope and observing wavelength. Accordingly, over a range in target redshift, a wide range in linear scales will be probed by the observations. In case of the \cite{2008ApJ...688..826L} sample, the linear scales probed range from $\sim0.4-8.5$kpc/\arcsec. 
Contamination by host galaxy emission or extended jet structures makes comparison between individual works difficult.
Despite a quite low angular resolution, in the X-ray domain, the high luminosity can generally be solely ascribed to the AGN \citep[cf., discussion in][]{2005ApJ...631..707P}. 
Galaxy clusters with significant extended X-ray emission present an exception. Here, high angular resolution and X-ray spectral information are necessary.

In general, $R_X$ and $R^*$ are well correlated and can, apparently, both be used to assess the radio loudness of AGN.

\subsubsection{Eddington Ratio}
\begin{figure*}[ht]
\centering
\includegraphics[width=17cm]{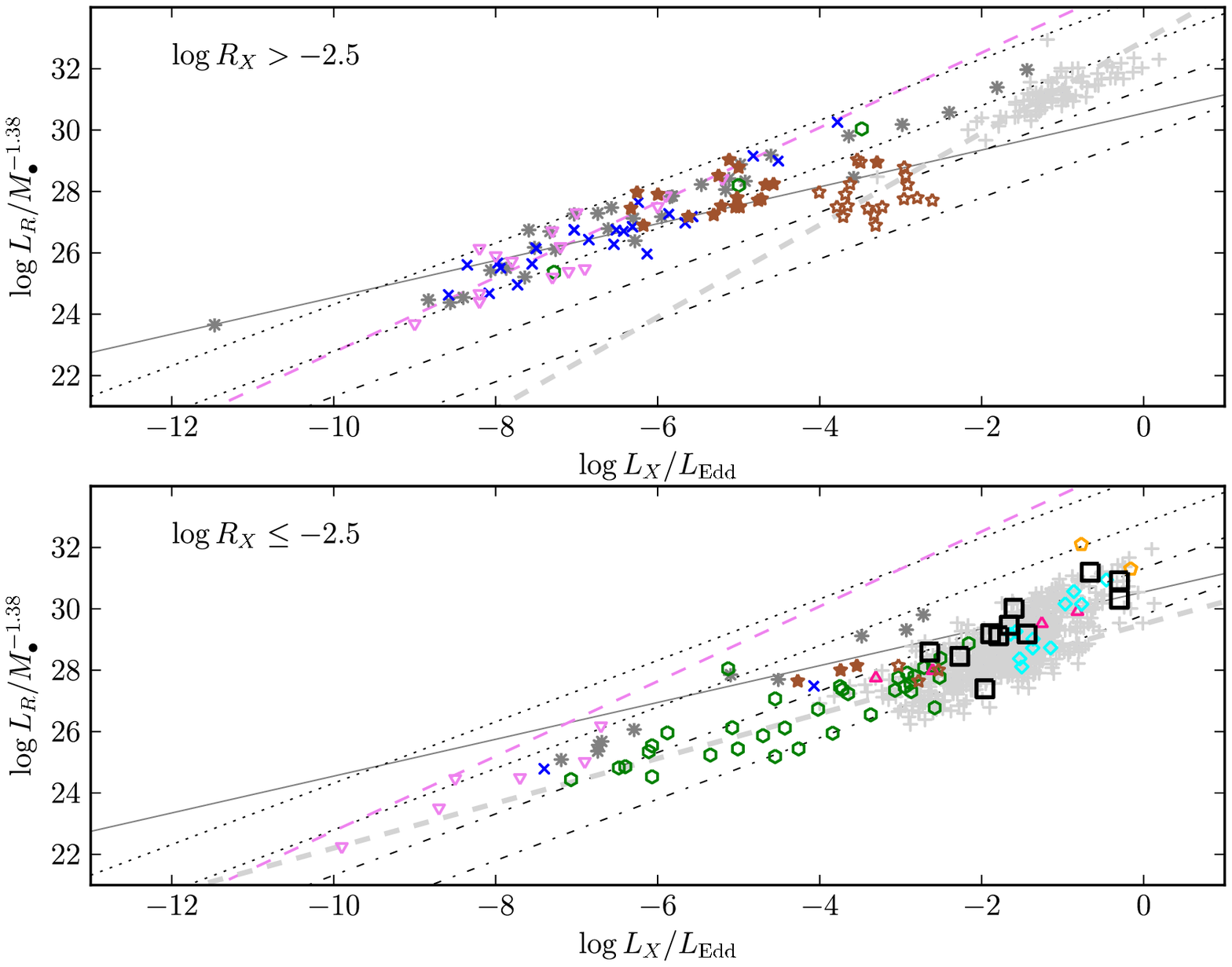}
\caption{
	$M_\bullet$ weighted 5~GHz radio luminosity versus the soft X-ray luminosity in units of the Eddington luminosity \citep[from the fundamental plane equation in][]{2003MNRAS.345.1057M}.
	Symbol shapes as in Fig. \ref{fig:RXRO}. 
	For clarity, our data are presented as open black squares.
	In addition, we show the data of \citet[grey asterisks]{2003MNRAS.345.1057M} and the low-luminosity AGN of \citet[violet, downright triangles]{2009ApJ...703.1034Y}.
	The dashed violet line represents to the fit of \cite{2009ApJ...703.1034Y} adopted to soft X-rays.
	The thin grey line shows the fundamental plane of \cite{2003MNRAS.345.1057M} adopted for the soft X-rays.	
	The thin black dotted (dot-dashed) lines indicate constant $R_X=-1.5$ ($R_X=-4.5$) for typical $\log M_\bullet[M_\odot]=6$ and 10 (upper and lower line), respectively. These lines have a slope of unity.
	Furthermore, we present data from \citet[stars]{2009MNRAS.396.1929H}, for which the fraction of accretion related X-ray emission could be determined. The brown colored full stars correspond to the full X-ray and the brown open stars to the accretion related X-ray luminosities.
	{\it Upper panel:} Presented are RL objects, $R_X>-2.5$, according to \cite{panessa_x-ray_2007}. The thick, dashed light grey line represents the fit of \cite{2008ApJ...688..826L} to their RL BLAGN.
	{\it Lower panel:} Presented are RQ objects, $R_X\leq -2.5$, according to \cite{panessa_x-ray_2007}. The thick, dashed light grey line represents the fit of \cite{2008ApJ...688..826L} to their RQ BLAGN.
	See text for details.
\label{fig:merloni}}
\end{figure*}

Assuming spherical accretion, for given black hole mass, the Eddington luminosity $L_\mathrm{Edd}$ calculates as
$L_\mathrm{Edd}\approx 1.26\times 10^{38} M_\bullet/M_\odot\mathrm{erg~ s}^{-1}.$
The Eddington ratio $\eta=L_\mathrm{bol}/L_\mathrm{Edd}$, where $L_\mathrm{bol}$ is the bolometric luminosity, can be calculated with an estimate of the bolometric luminosity based on the rest-frame 5100\AA\ luminosity density \citep{2004ApJ...601..676V}\footnote{Actually, the optical luminosity density used in \cite{2004ApJ...601..676V} is measured at 4400\AA. Arguably, the commonly used luminosity density at 5100\AA~is sufficiently close in wavelength to yield consistent values.}:
$L_\mathrm{bol}\approx 9.47\times \lambda L_\lambda(5100\AA)\,\mathrm{erg~s}^{-1}.$
The bolometric luminosity of AGN is proportional to the accretion rate $L_\mathrm{bol}\propto\dot M_\bullet c^2$, therefore $\eta\propto\dot M_\bullet/\dot M_\mathrm{Edd}$ is an indirect measure of the accretion rate relative to the Eddington rate.
	Accretion scenarios are, generally, more complex than the spherical case. 
	\cite{2011arXiv1112.5483B}, for example, carry out simulations of non-spherical accretion onto an SMBH and find deviations of spherical accretion as a function of the X-ray luminosity. In units of the Eddington luminosity, they find that for low $L_X\approx 0.005$, gas accretes in a spherical fashion, whereas, for intermediate $L_X\approx 0.02$, the flow develops prominent non-spherical features. The latter being a mixture of cold gas clumps falling towards the SMBH, forming a filamentary structure and hot gas being slowed down or even driven outwards again due to buoyancy.
	The objects presented here cover the whole range of $L_X$ values.
	Our targets, as well as the BLAGN have $L_X\sim 0.2$ (cf., Fig. \ref{fig:merloni} and discussion in Sect. \ref{sec:fp}). 
	At face value, the Eddington ratio has to be taken with care and provides only a coarse impression of the accretion efficiency. 
For our sample, we find a range of Eddington ratios.
Table \ref{tab:eddington} lists the Eddington ratios for the 11 objects with SMBH mass estimates, together with the ratio of soft X-ray over Eddington luminosity.
On average $\eta=0.18$. The NLS1s have the highest Eddington ratios, $\eta\sim 0.45$, whereas the two LINERs and the RL QSO (515) have the lowest ratios, $\eta\sim 0.06$. 
These values are typical for luminous AGN \citep[e.g.,][]{2007ApJ...667..131G}. 
Here, $\eta$ for the LINERs is on the high end of values found \citep[e.g.,][]{2005ApJ...620..113D}, whereas $\eta$ for the NLS1s is on the low end of values found for this class \citep[e.g.,][]{2004MNRAS.352..823B}. 
We recognize a trend of decreasing $\eta$ with increasing $M_\bullet$, which is commonly found for AGN \citep{2007ApJ...667..131G}.

For estimating the bolometric luminosities, we have fitted the rest-frame shifted SDSS spectra with a linear combination of a power law and two stellar population templates (young and old), including reddening in all components, to get an impression of a possible host galaxy contribution \citep[cf.,][]{diss07}. For the SY1s and NLS1s we find no significant host galaxy contributions in the spectra and therefore extract the foreground extinction corrected 5100\AA~ fluxes directly from the spectra.
For the two LINER sources we have used our host-subtracted estimates of the 5100\AA~luminosity density. 
It has to be noted that the bolometric correction applied as described above is, strictly speaking, valid only in a statistical sense, since the bolometric luminosity strongly depends on the shape of the spectral energy distribution of the individual objects \citep[cf.,][]{2008ARA&A..46..475H}.

\begin{table}
	\begin{center}
		\caption{Eddington parameters ordered by spectral class.\label{tab:eddington}}
		\begin{tabular}{ccccc}
		\hline
		\hline
		ID & spectral class & $\log L_\mathrm{Edd}/$erg~s$^{-1}$ & $\eta$ & $\log L_{X,\mathrm{rest}}/L_\mathrm{Edd}$\\
		\hline
		1  & NLS1  & 44.6 & 1.49 &  -0.66 \\
		2  & NLS1  & 44.8 & 0.52 &  -1.60 \\
		3  & NLS1  & 46.1 & 0.22 &  -1.79 \\
		7  & NLS1  & 45.4 & 0.18 &  -1.89 \\
		8  & NLS1  & 44.9 & 0.14 &  -1.44 \\
		6  & SY1   & 45.5 & 0.31 &  -2.27 \\
		10 & SY1   & 46.0 & 0.14 &  -1.66 \\
		11 & SY1   & 47.5 & 0.01 &  -2.64 \\
		13 & SY1   & 45.7 & 0.26 &  -1.94 \\
		4  & LINER & 45.6 & 0.15 &  -0.30 \\
		5  & LINER & 46.7 & 0.03 &  -1.96 \\
		\hline
	\end{tabular}
\end{center}
\end{table}


\subsubsection{The fundamental plane of black hole activity}
\label{sec:fp}
It has been recognized that stellar-mass black holes and supermassive black holes share a common {\it black hole fundamental plane} (FP), spanned by the three parameters X-ray luminosity, radio luminosity, and black hole mass \citep[e.g.,][]{2003MNRAS.344...60G,2003MNRAS.345.1057M,2004A&A...414..895F}: $\log L_R=\zeta_{RX}\log L_X+\zeta_{RM}\log M_\bullet+C$. 
This relation (i.e., its correlation coefficients) implies a fundamental connection between radio and X-ray emission producing processes. 
The \cite{2003MNRAS.345.1057M} fit gives $\zeta_{RX}=0.6\pm 0.11$, $\zeta_{RM}=0.78^{+0.11}_{-0.09}$, and $C=7.33^{+4.05}_{-4.07}$. 
Currently, the FP is interpreted in the context of synchrotron emission from a jet \citep[both in the radio and X-ray regime, e.g.,][]{2004A&A...414..895F,2009ApJ...703.1034Y} or an advection-dominated accretion flow producing X-rays together with a jet visible in the radio \citep{2003MNRAS.345.1057M}.
There are, however, complications related to selection and observational effects \citep[cf.,][]{2006A&A...456..439K}. 

In Fig. \ref{fig:merloni}, according to the visualization by \citet[their Fig. 7]{2003MNRAS.345.1057M}, we show the black-hole mass weighted 5~GHz luminosities versus soft X-ray luminosities in units of Eddington luminosity. 
The two panels correspond to RL ($\log R_X>-2.5$) and RQ ($\log R_X\leq -2.5$) AGN.
Our measurements are presented as black open squares in the RQ regime and the data points follow the Merloni FP relation (thin grey line, slope $\zeta_{RX}=0.6$) closely.
For comparison, we include the same data sets as for Fig. \ref{fig:RXRO}.
Additionally, we show data for low-luminosity AGN (LLAGN) of \citet[violet downright triangles]{2009ApJ...703.1034Y} and the data set of \citet[grey asterisks]{2003MNRAS.345.1057M}.
The fit to the \citet[dashed, thin violet line]{2009ApJ...703.1034Y} LLAGN has slope $\zeta_{RX}=1.23$\footnote{\cite{2009ApJ...703.1034Y} find the following correlation coefficients: $\zeta_{RX}=1.22$, $\zeta_{RM}=0.23$, $C=-12.46$}. 
The thin dotted and dash-dotted grey lines correspond to constant $R_X=-1.5$ and $R_X=-4.5$, respectively. 
The upper and lower lines of each pair (dotted, dash-dotted) correspond to black hole masses $M_\bullet=10^6M_\odot$ and $M_\bullet=10^{10}M_\odot$, respectively. 
\cite{2006ApJ...645..890W} and \cite{2008ApJ...688..826L} find different correlation slopes for RL ($\zeta_{RX}=1.5\pm 0.08$, $\zeta_{RM}=-0.2\pm 0.1$, $C=0.05\pm0.1$) and RQ ($\zeta_{RX}=0.73\pm 0.10$, $\zeta_{RM}=0.31\pm 0.12$, $C=-0.68\pm0.07$) BLAGN, which are individually different from the relation found by \cite{2003MNRAS.345.1057M}.
We plot their correlations (RL, RQ) as thick dashed lines into the corresponding panel in Fig. \ref{fig:merloni}.
The dependence on the black hole mass in their large sample is weak and only for RQ BLAGN with black hole masses $\lesssim 10^{7.3}M_\odot$ a clear positive trend is observed. 
\cite{2006A&A...456..439K}, e.g., show that the Eddington ratio is an important ingredient, with strongly sub-Eddington objects following the FP relation much tighter.
\cite{2009MNRAS.396.1929H} demonstrate the importance of a proper separation of accretion-related from jet-related X-ray emission. Based on X-ray spectral modeling, they found that the \cite{2003MNRAS.345.1057M} relation lies much closer to the accretion-related X-ray luminosities of their sample.
We demonstrate this finding by plotting their data, for which such a decomposition is possible. Brown filled stars represent the total X-ray luminosities, whereas brown open stars represent the corresponding accretion related X-ray luminosities. 
It appears that the current FP reflects a relation between accretion power and jet power, rather than the structural properties of (relativistic) jets \citep{2009MNRAS.396.1929H}.
Especially, low angular resolution, broad band X-ray data are prone to mixing effects (i.e. corona, jet, warm absorber, etc.), which complicates the interpretation. This is also true for our sample, which is based on the soft X-ray RASS data.
Nevertheless, \cite{2008ApJ...688..826L} study the relation between soft X-rays and the broad H$\beta$ emission for their sample. They find a strong correlation and interpret it as a signature of the accretion process. Besides the different normalization, the slope of their correlation for RQ BLAGN is quite similar to that of \cite{2003MNRAS.345.1057M}. At least, the RQ BLAGN appear to be accretion dominated.

%
As \cite{2009MNRAS.396.1929H} and \cite{2009ApJ...703.1034Y} show, a proper decomposition of the X-rays into jet- and accretion-related components introduces considerably different distributions in diagrams like Fig. \ref{fig:merloni}. 
As soon as the sample chosen covers only a narrow range in radio loudness, then this results in a 'correlation' slope of unity. This is indicated by lines of $\log R_X=constant$ in the Figure.
At this point, without a proper accretion/jet decomposition, the FP does not appear to provide new information about the radio/X-ray connection.
There is certain room for scatter, as indicated by the dotted and dot-dashed lines (dotted for $\log R_X=-1.5$ and dashed for $\log R_X=-4.5$), the upper ones corresponding to $M_\bullet=10^6M_\odot$ and the lower ones to $M_\bullet=10^{10}M_\odot$. 


Similarly, as radio emission is produced by relativistic jets, Doppler boosting of the synchrotron radiation could significantly affect the observed radio flux densities, resulting in a vertical shift in the FP diagram.
\cite{2008ApJ...688..826L} investigate the influence of relativistic beaming on their BLAGN sample. They estimate the intrinsic (unbeamed) radio power with the observed X-ray luminosity through the radio/X-ray correlation derived from RQ AGN only (their Fig. 13). 
Typically, BLAGN have boosting factors less than 30, whereas in the SDSS-based BLAGN sample, factors of up 1000 (which can only produced in BL Lac objects) are found. 
Nevertheless, for the majority of their sample having factors of $\lesssim 100$, boosting alone is not sufficient to explain the observed larger radio powers of RL BLAGN \citep[cf.,][]{2004MNRAS.355L...1H}.
Though, as discussed above, using soft X-ray luminosities of the RQ AGN might introduce uncertainties in the estimation of beaming factors. 
All this points towards a direction in which it is not straightforward to read of the distribution of data points the connection between X-ray and radio emission for given samples. In other words, the details of the sample selection and observation are essential. 

Interestingly, for RQ AGN, at $\log L_X/L_\mathrm{Edd}\gtrsim -1$ there appears to be a limit in the Eddington-weighted X-ray luminosity in Fig. \ref{fig:merloni}. 
In this region, objects tend to have increased radio luminosities. This is clearly visible for the \cite{2008ApJ...688..826L} BLAGN. 
As discussed by the authors, the soft X-ray luminosity likely reflects the accretion process, because of the tight correlation with the broad Balmer line luminosities. 
This could mean that the SMBHs are accreting most efficiently around their Eddington limit and the increased radio luminosity can then be attributed to (beamed) jet emission. 
Our higher angular resolution radio data together with the ROSAT data follow the same trend\footnote{This takes account of the similar database (i.e., SDSS/ROSAT/FIRST matches)}. Nevertheless, we do not find indications for kpc-scale jets and we would require still higher angular resolution radio observations in order to associate beamed radio emission with high brightness temperatures \citep[cf.,][]{2007PASJ...59..703D}.

\section{Summary}
\label{sec:summary}
We have presented 18~cm MERLIN observations of 13 intermediate redshift X-ray bright AGN, four of which have also been studied at 18~cm with Western EVN.
This work has been intended to be a pilot study to investigate the energetic components (i.e., starburst vs. active nucleus) in the nuclei of such AGN at high angular resolution.

Despite the variety of spectroscopic types (SY1, NLS1, LINER, BL Lac, and passive), the 18~cm emission is essentially compact in both MERLIN (on $\sim 500$pc scales) and EVN maps (on $\sim 40$~pc scales), with up to 50\% of the lower-resolution VLA flux densities being resolved out by our interferometric observations. 
There are no indications for distinctive jets, except for a BL Lac object (9) and an FR-II QSO (object 11). 

Based on various SFR indicators, star formation does not play a dominant role, and the compact radio emission is, therefore, most likely related to the AGN. 
However, we cannot discern between accretion or jet related radio emission at this point, because of the lack of spatial and spectral resolution both in the radio and X-ray domain. 

Using the soft X-ray and 5~GHz luminosities, we placed our sample objects on the fundamental plane of black hole activity diagram (Fig. \ref{fig:merloni}). 
Our targets follow the general trends seen in the diagram of RQ AGN. 
Here we used the radio over X-ray luminosity ratio as an indicator for the radio loudness.
Generally, the slope of the FP is related to a specific accretion/jet model. 
As discussed by several authors \citep[e.g.,][]{2009ApJ...703.1034Y, 2009MNRAS.396.1929H, 2006A&A...456..439K}, care has to taken when estimating accretion/jet related X-ray and radio fluxes for individual objects based on the FP. 
This, on the other hand, suggests that there is no simple interpretation of the FP without a proper disentangling of the accretion/jet components and consideration of object types (e.g., QSO, LLAGN, radio galaxy). 

Our data lie close to the relation found by \cite{2003MNRAS.345.1057M}, which describes properties of the jets rather than the accretion process.
Higher angular resolution radio images at several frequencies, as well as hard X-ray spectra are required to study the emission processes in our sources in more detail and to better constrain their position in the fundamental plane diagram.

The {\it e}MERLIN upgrade will allow for significantly improved sensitivity for imaging (down to a noise level of a few $\mu$Jy) and, correspondingly, to an increased sensitivity for extended emission. 
	On VLBI level, the e-EVN project will provide similar improvements, but for much higher angular resolution. 
Such advancements are instrumental to further the characterization of the relevant emission components that might determine the location of AGN in the FP.

\begin{acknowledgements}
The authors kindly thank the anonymous referee for fruitful comments and suggestions.
The authors kindly thank Anita Richards, Tom Muxlow, and Giuseppe Cimo for their help during data reduction, as well as Martin Hardcastle for kindly providing additional information on their X-ray study of 3CRR sources. 
This work has benefited from research funding from the European Community's sixth Framework Programme under RadioNet R113CT 2003 5058187. MERLIN is a National Facility operated by the University of Manchester at Jodrell Bank Observatory on behalf of STFC. The European VLBI Network is a joint facility of European, Chinese, South African and other radio astronomy institutes funded by their national research councils. Part of this work was supported by the German
      \emph{Deut\-sche For\-schungs\-ge\-mein\-schaft, DFG\/} project
      numbers SFB~494 and SFB~956. JZ acknowledges support from the European project EuroVO DCA under the Research e-Infrastructures area (RI031675).
\end{acknowledgements}

\bibliographystyle{aa}
\bibliography{merlin_evn}




\FloatBarrier
\appendix
\section{Notes on individual targets}
\label{sec:notes}


\paragraph{SDSS J153911.17+002600.7 (1):}
This AGN is a NLS1 galaxy \citep{2006ApJS..166..128Z,2002AJ....124.3042W}. From its RASS observation, a hardness ratio of $HRI=0.46\pm0.45$ and a very soft photon index $\Gamma=2.8\pm 0.9$ have been deduced by \cite{2002AJ....124.3042W}. Its SDSS images exhibit some irregular structures around the nucleus. Part of this structure is a foreground star. The remaining faint features might belong to the host galaxy, or to some background galaxy.
It is the radio loudest AGN in our sample, when neglecting the two jet-dominated sources (9 and 11; see below).

\paragraph{SDSS J150521.92+014149.8 (2):}
This NLS1 galaxy shows faint signatures of a spiral host galaxy. It also exhibits a strongly blueshifted [OIII] $\lambda5007$ emission line (by about 165km/s), which is commonly found in NLS1s and might be related to outflows or inflows \citep{2005AJ....130..381B}.

\paragraph{SDSS J134206.57+050523.8 (3):}
This is an NLS1 \cite{2001ApJ...546..744N, 2006ApJS..166..128Z} and is also a member of the \cite{2008ApJ...688..826L} sample.

\paragraph{SDSS J150407.51-024816.5 (4):}
This elliptical LINER galaxy is the central galaxy of the compact X-ray luminous galaxy cluster \object{RXC J1504-0248} with one of the most massive nearby cooling cores \citep{2005ApJ...633..148B}. 
The black hole mass, estimated from the [SII] line width \citep[cf.,][]{2007ApJ...667L..33K}, of about $3\times 10^7\,M_\odot$, is somewhat less than those estimated for the brightest cluster central galaxies \citep[$10^{8-9}M_\odot$, e.g.,][]{2006ApJ...652..216R,2004ApJ...612..797F}.

\begin{figure}[ht]
\begin{center}
\resizebox{\hsize}{!}{\includegraphics{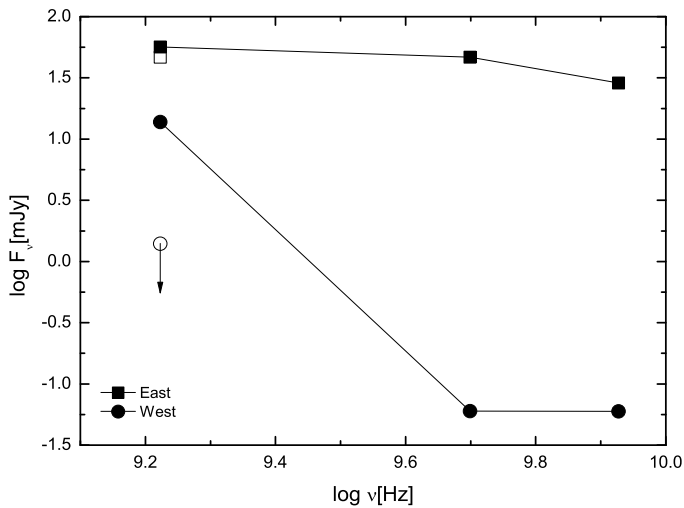}}
\end{center}
\caption{Target 4 radio continuum spectra of eastern and western components: MERLIN at 1.67~GHz (open symbols represent EVN measurements), EVN at 5~GHz, and VLA at 8.46~GHz.}
\label{fig:127_radio_spectra}
\end{figure}

In our MERLIN 18~cm images we detected two sources (4E[ast] and 4W[est]) at the optical position of the galaxy (Fig. \ref{fig:maps_evn}). The two components are separated by about 0\farcs 5, which corresponds to 1750~pc at $z=0.217$.
In the VLA archive, we found a 3.5~cm data set at an angular resolution ($\approx 0.25\arcsec$) similar to that of our MERLIN image.
Both components are detected at 3.5~cm, with $F_\mathrm{peak}(\mathrm{E})=28.78\pm 0.07$~mJy and $F_\mathrm{peak}(\mathrm{W})=0.60\pm 0.07$~mJy.
The EVN 6~cm image recovers both components, whereas in the EVN 18~cm image, we only detect 4E. The 18~cm limiting peak flux density of 4W is 1.4~mJy. 
Figure \ref{fig:127_radio_spectra} displays the continuum spectra of 4E and 4W.
The MERLIN/VLA two-point spectral index $\alpha^{18\mathrm{cm}}_{3.5\mathrm{cm}}=-1.419\times \log(S_{3.5\mathrm{cm}}/S_{18\mathrm{cm}})$ is 0.42 for 4E and 2.05 for 4W.
The EVN two-point spectral index $\alpha^{18\mathrm{cm}}_{6\mathrm{cm}}=-2.0959\times \log(S_{6\mathrm{cm}}/S_{18\mathrm{cm}})$ is 0.14 for 4E and $<0.58$ for 4W.
The eastern component has a relatively flat 18-6-3.5~cm spectrum. 
The western component is steeper, but flattens towards shorter wavelengths. 
The flat nuclear spectrum might be the result of a set of closely spaced homogeneous synchrotron components or optically thin bremsstrahlung in a disk wind \citep[cf.,][]{blundell_origin_2007}. 
The SDSS image of the host galaxy is regular and shows no signs of a recent merger. Together with the steep spectral index, the western component does not appear to be a companion nucleus, but rather a jet component. This is consistent with an asymmetry seen in the central X-ray brightness distribution \citep[Fig. 9 in][]{2005ApJ...633..148B}, which might be the result of interaction between the radio jet and the hot cluster medium. 
Further conclusion are limited due to the fact that the EVN 18~cm data were hampered by bad weather. Further and deeper radio data are required to study the impact of the potential radio jet onto the surrounding environment. 

\paragraph{SDSS J010649.39+010322.4 (5):}
This LINER galaxy lies at the heart of the galaxy cluster \object{RXC J0106.8+0103} or \object{Zw 348} \citep[e.g.,][]{2000MNRAS.318..333E,2000ApJS..129..435B}. In this cluster, the AGN is expected to be a major contaminant of the overall X-ray emission \citep{2000ApJS..129..435B}. 
However, from the small $R_X$ and $R^*$ values in Fig. \ref{fig:RXRO}, it can be seen that the X-rays still appear to be dominated by the cluster emission.
The FIRST image reveals a slight extended emission component to the southeast in addition to the compact core (the extend parameter $\theta_\mathrm{FIRST}\approx 1.3$), which, being barely above the noise level, could be an indication for a jet component. Our MERLIN image does not reveal any extended emission, since it might resolve out faint extended emission.
This galaxy is also luminous at UV wavelengths (based on GALEX data). It lies on the boundary of between large UV luminous galaxies (UVLGs) and compact UVLGs as defined by \cite{2007ApJS..173..441H}.
Large UVLGs are principally normal galaxies, but are very luminous primarily because of their large size. Compact UVLGs, on the other hand, are low mass, relatively metal poor systems that often have disturbed morphology. These systems are characterized by intense ongoing star formation.
\cite{2007ApJS..173..441H} estimate the SFR by SED fitting. 
$SFR_{UV}\approx 2.4$ is compatible with our FIRST estimated SFR (Tab. \ref{tab:derived}). Our MERLIN results show a much smaller SFR, which indicates that the star formation is significantly extended throughout the host and is resolved out by the interferometer. The increasing brightness temperature at MERLIN angular resolution indicates for an AGN/jet origin of the unresolved emission.

\paragraph{SDSS J074738.38+245637.3 (6):}
This early-type Seyfert 1 has a companion \object{SDSS J074738.25+245626.1} at $z=0.131$. Low surface-brightness features in the SDSS images are located between the two galaxies. Gravitational interaction could therefore be responsible for the SFR of $\sim 50M_\odot$~yr$^{-1}$ estimated from the large aperture data. 
MERLIN appears to resolve out a fraction of about 80\% of the FIRST emission.
The resulting SFR estimate is comparable to the [OII] one, which is about $10M_\odot$~yr$^{-1}$.
This is the most radio quiet source of the sample, and with the MERLIN measurement it becomes even more radio quiet, $\log R^*_\mathrm{MERLIN}\approx -0.2$.
The X-ray luminosity is typical for Seyfert 1 galaxies, which in turn indicates that only a small fraction of the MERLIN flux density is related to star formation, in order to follow the X-ray/radio correlation. 
MERLIN probes galactic scales of $\sim 450$ pc, whereas the 3\arcsec diameter SDSS fiber probes a region of about 7~kpc diameter. 
This suggests that the [OII] luminosity might be related to the extended, resolved out FIRST radio emission. 
Higher angular resolution radio observations are required to resolve this issue.


\paragraph{SDSS J030639.57+000343.1 (7):}
This NLS1 has a spiral morphology. The galaxy is possibly a member of a group, with a projected distance of about 110~kpc to the nearest neighbor. It has a redshifted [OIII] $\lambda5007$ emission line ($\sim46$km s$^{-1}$), which, as discussed above, indicates for a complex narrow-line region kinematics \citep{2005AJ....130..381B}.

\paragraph{SDSS J081026.07+234156.1 (8):}
\cite{2006AJ....131.1948W} classified this AGN as an NLS1 galaxy. Its optical appearance is compact. 
It is the radio faintest object in our sample. With the MERLIN measurement, it becomes the second most radio quiet AGN, $\log R^*_\mathrm{MERLIN}\approx 0.3$.

\paragraph{SDSS J162332.27+284128.7 (9):} This optically unresolved object has been classified as a BL Lac object \citep{2008AJ....135.2453P}. Due to the absence of spectral features, no black hole mass could be estimated.
This might also be a problem for its redshift, which is taken from the SDSS database.
This object is not particularly radio loud (Fig. \ref{fig:RXRO}), though we are supposed to observe the jet almost face on.

\paragraph{SDSS J080322.48+433307.1 (10):}
This SY1 galaxy shows some extremely redshifted [OIII] emission \citep[$\sim 57$km\ s$^{-1}$,][]{2005AJ....130..381B} and is also included in the large catalogs by \cite{2006ApJ...645..890W} and \cite{2008ApJ...688..826L}. Due to a nearby guide star (\object{SDSS J080324.87+433330.1}), it is also suitable for AO-assisted observations.

\paragraph{SDSS J080644.41+484149.2 (11):} 
This source is the only FRII RL QSO \citep{2006AJ....131..666D} in our sample, having a nearby natural guide star (\object{SDSS J080645.63+484156.6}). Its optical appearance is unresolved.
Although being RL in the classical sense, the $R_X$ measurement would still result in a RQ AGN. We have only used the compact nuclear radio emission. This could mean that the 40pc scale emission not that strongly influenced by the jet component. 

\paragraph{SDSS J092710.60+532731.6 (12):}
This radio galaxy shows a passive, elliptical optical spectrum. It has a nearby natural guide star (\object{SDSS J092707.09+532742.3}) and is likely the brightest cluster galaxy in the \object{Zw 2379} galaxy cluster \citep[$z=0.205$;][]{2008MNRAS.384.1502S}. 
Its high EVN brightness temperature underlines the AGN nature of the radio emission. 

\paragraph{SDSS J134420.87+663717.6 (13):}
IRAS~13428+6652 has been identified as Seyfert 1.5 by \cite{1996ApJS..106..341M}. It appears to be a major merger with an extended tidal tail and possibly two nuclei. It is suitable for AO assisted studies (\object{SDSS J134427.10+663705.2}). The source has been detected at 20~cm by NVSS. However, we could not detect any compact flux density at 18~cm with MERLIN. Since it is an IR luminous galaxy ($L_\mathrm{FIR}\approx 5\times 10^{11}L_\odot$), the radio emission is most likely dominated by extended star formation.

\end{document}